\documentclass[12pt]{article}
\usepackage{graphicx}
\usepackage{amsmath}
\usepackage{amssymb}
\usepackage{caption2}
\begin{document}
\bibliographystyle{prsty}
\begin{center}
{\large {\bf \sc{  Analysis of the Isgur-Wise function of the
$\Lambda_b \to \Lambda_c$ transition   with light-cone QCD sum rules  }}} \\[2mm]
Zhi-Gang Wang \footnote{ E-mail,wangzgyiti@yahoo.com.cn.  }    \\
 Department of Physics, North China Electric Power University, Baoding 071003, P. R. China
\end{center}

\begin{abstract}
In this article, we use the light-cone QCD sum rules to relate the
$\Lambda_b$ baryon light-cone distribution amplitudes to the
Isgur-Wise function $\xi(\omega)$ of the $\Lambda_b \to \Lambda_c$
transition, and obtain a simple relation. The numerical value of the
Isgur-Wise function $\xi(\omega)$ is consistent with the prediction
of the QCD sum rules.
\end{abstract}

PACS numbers:  12.38.Lg; 14.20.Mr, 14.20.Lq

{\bf{Key Words:}}  $\Lambda_Q$ baryon, Isgur-Wise function,
Light-cone QCD sum rules
\section{Introduction}
The semileptonic  decay $b\to c$  is an important process in
extracting the CKM matrix element  $V_{cb}$ and serves  as  a
laboratory for studying the nonperturbative  QCD effects. In the
baryon sector,  the semileptonic decay
$\Lambda_b\to\Lambda_c\ell\bar\nu_\ell$, which takes  place through
the process $b\to cW \to c\ell\bar\nu_\ell$ at the quark level, has
 attracted  much attention.
 The charm and bottom baryons
(e.g. $\Lambda_c$ and $\Lambda_b$) which contain a heavy quark and
two light quarks are particularly interesting for studying dynamics
of the light quarks in the presence  of a heavy quark. They behave
as the QCD analogue of the familiar hydrogen bounded by the
electromagnetic interaction, and serve as an excellent ground for
testing predictions of the constituent  quark models and heavy quark
symmetry \cite{Review,Heavybook}.

In the heavy quark limit, we can express the hadronic form-factor
$\Lambda_b\to\Lambda_c$  in terms of the Isgur-Wise function $\xi
(\omega)$ \cite{Heavybook},
\begin{equation}
\langle\Lambda_c(v')|\bar{c}_{v'}(0)\Gamma
b_{v}(0)|\Lambda_b(v)\rangle = \xi
(\omega)\overline{U}_{\Lambda_c}(v')\Gamma U_{\Lambda_b}(v)\,,
\end{equation}
where the $v$ and $v'$ are  velocities of the heavy quarks $b$ and
$c$ respectively, $\omega=v\cdot v^\prime$, and the $U(v)$ is the
Dirac spinor. The Isgur-Wise function $\xi(\omega)$ is normalized to
1 at   zero recoil $\omega=1$. In the weak decay $b\to cW$,  the
light degrees of freedom undergo a corresponding transition due to
the gluon exchanges with the heavy quarks, and the Isgur-Wise
function $\xi(\omega)$ has copious information about the dynamics of
the light degrees of freedom. The physical region of the $\omega$
($=\frac{M_{\Lambda_b}^2+M_{\Lambda_c}^2-q^2}{2M_{\Lambda_b}M_{\Lambda_c}}$)
is rather small, about $1.0-1.43$. The existing theoretical
estimations of the slope parameter $\rho^2$   vary from $0.5$ to
$1.5$, one can consult  Ref.\cite{HuangMQ05} for more literatures.
Using an exponential parametrization
$\xi(\omega)=\exp\left[-\rho^2(\omega-1)\right]$, the DELPHI
Collaboration  obtain a value  $\rho^2=1.59\pm 1.10 $; after taking
into account the observed event rates and  adding  the normalization
condition  $\xi(1)=1$, they reach the value $\rho^2=2.03\pm
0.46^{+0.72}_{-1.00}$ \cite{DELPHI2004}, the uncertainty is very
large.

In Refs.\cite{Khodjamirian05,KhodjamirianB07},  Khodjamirian et al
derive  new sum rules for the $B\to \pi,K ,  \rho,K^*$ form-factors
from the correlation functions expanded near the light-cone in terms
of the $B$-meson distribution amplitudes, and suggest QCD sum rules
motivated models for the  three-particle $B$-meson light-cone
distribution amplitudes, which satisfy the  exact relations between
the two-particle and three-particle $B$-meson light-cone
distribution amplitudes \cite{Qiao2001}. The $B$-meson light-cone
QCD sum rules have been applied to  the form-factors  $B \to
a_1(1260)$ \cite{Wang08} and $B \to D, D^*$ \cite{Khodjamirian09}.
In Ref.\cite{DeFazioJHEP},  De Fazio et al study the sum rules for
the heavy-to-light transition form-factors at large recoil derived
from the correlation functions with interpolating currents for the
light pseudoscalar (or vector) fields in soft-collinear effective
theory and the $B$-meson light-cone distribution amplitudes.

In Ref.\cite{Ball2008}, Ball et al  perform  a complete
classification of the three-quark distribution amplitudes of the
$\Lambda_b$ baryon in QCD in the heavy quark limit and  discuss the
relevant  features, and  derive a renormalization-group equation
which governs the scale-dependence of the leading-twist light-cone
distribution amplitudes. Furthermore, they suggest simple models of
the light-cone distribution amplitudes and estimate the relevant
parameters based on  the first few moments using the QCD sum rules.

In this article, we study the Isgur-Wise function $\xi(\omega)$ of
the transition $\Lambda_b \to \Lambda_c$ with the $\Lambda_b$-baryon
light-cone QCD sum rules, i.e. we use the light-cone sum rules  to
relate the $\Lambda_b$-baryon light-cone distribution amplitudes to
the Isgur-Wise function $\xi(\omega)$.

The article is arranged as: in Section 2, we derive the Isgur-Wise
function $\xi(\omega)$ with the light-cone QCD sum rules; in Section
3, the numerical result and discussion; and in Section 4 is reserved
for conclusion.

\section{ Isgur-Wise function $\xi(\omega)$ with light-cone QCD sum rules}

 We study the Isgur-Wise function $\xi(\omega)$ with the
 two-point correlation functions $\Pi^i_{\mu}(p,q)$,
\begin{eqnarray}
\Pi^i_{\mu}(p,q)&=&i \int d^4x \, e^{i p \cdot x} \langle 0
|T\left\{\eta_i(x) J_{\mu}(0)\right\}|\Lambda_b(v)\rangle \, ,
\end{eqnarray}
where
\begin{eqnarray}
\eta_1(x)&=&\epsilon_{ijk}u^i(x)C\gamma_5\not\!n d^j(x) c^k_{v}(x) \, ,\nonumber\\
\eta_2(x)&=& \epsilon_{ijk}u^i(x)C\gamma_5 d^j(x) c^k_{v}(x) \, ,\nonumber\\
 \eta_3(x)&=&\epsilon_{ijk}u^i(x)C\gamma_5 i  \sigma_{\bar n n} d^j(x) c^k_{v}(x) \,
 ,\nonumber\\
 J_\mu(x)&=&\bar {c}_{v'}(x)\gamma_\mu (1-\gamma_5) b_v(x)\, ,
 \end{eqnarray}
 the quark currents $\eta_i(x)$ ($i=1,2,3$) interpolate the heavy baryon
 $\Lambda_c$,  $\sigma_{\bar n n}=\sigma_{\mu\nu}\bar{n}^\mu n^\nu$, the $n_\mu$ and $\bar n_\mu$ are
 light-like vectors and the $C$ is the charge conjugation matrix.

Based on the assumption of the quark-hadron duality
\cite{SVZ79,Reinders85}, we insert  a complete set of intermediate
states with the same quantum numbers as the current operators
$\eta_i(x)$  into the correlation functions $\Pi^i_{\mu}(p,q) $ to
obtain the hadronic representation. After isolating the ground state
contribution from the pole term of the $\Lambda_c$ baryon,  the
correlation functions
 $\Pi^i_{\mu}(p,q)$  can be expressed in the following form,
\begin{eqnarray}
\Pi^i_{\mu}(p,q)&=&\frac{f^i_{\Lambda}}
  {\bar{\Lambda}-p\cdot v'}   \frac{1+\not\!v'}{2} \gamma_\mu  \xi(\omega)U(v)  + \cdots\, ,
 \end{eqnarray}
 where the $\bar{\Lambda}$ is the bound energy in the heavy quark limit, $\bar{\Lambda}=M_{\Lambda_b}-m_b=M_{\Lambda_c}-m_c$.
 We have used the standard definition  for the pole residues (or the coupling constants)  $f^i_{\Lambda}$,
 \begin{eqnarray}
 \epsilon_{ijk}\langle 0| \left[u^i(0)C\gamma_5 \not\!n d^j(0)\right] h^k_{v}(0)
  |\Lambda(v)\rangle &=& f^{2}_\Lambda  U(v)\,,
\nonumber\\
 \epsilon_{ijk}\langle 0| \left[u^i(0)C\gamma_5 d^j(0)\right] h^k_{v}(0)
  |\Lambda(v)\rangle &=& f^{1}_\Lambda  U(v)\,,
\nonumber\\
 \epsilon_{ijk}\langle 0| \left[u^i(0)C\gamma_5
 i  \sigma_{\bar n n} d^j(0)\right] h^k_{v}(0)
  |\Lambda(v)\rangle &=&  f^{3}_\Lambda  U(v)\, ,
\end{eqnarray}
and $f^{3}_\Lambda=2f^{1}_\Lambda$.

 In the following, we briefly outline
the operator product expansion for the correlation functions
$\Pi^i_\mu(p,q)$ in perturbative QCD. The calculations are performed
at the large space-like momentum region $|p^2|\gg \Lambda_{QCD}$ and
$ q^2\gg \Lambda_{QCD}$. We write down the propagator of the $c$
quark and the light-cone distribution amplitudes of the $\Lambda_b$
baryon in the heavy quark limit,
\begin{eqnarray}
\langle 0 | T \{h_v(x)\, \bar{h}_v(0)\}| 0 \rangle
&=&\frac{1+\not\!v}{2}
 \int_0^\infty d\lambda \delta^4(x-v\lambda) \, ,
 \end{eqnarray}
\begin{eqnarray}
 \epsilon_{ijk}\langle 0| \left[u^i(t_1 n)C\gamma_5 \not\!n d^j(t_2 n)\right] h^k_{v}(0)
  |\Lambda_b(v)\rangle &=& f^{2}_\Lambda \Psi_2(t_1,t_2) U(v)\,,
\nonumber\\
 \epsilon_{ijk}\langle 0| \left[u^i(t_1 n)C\gamma_5 d^j(t_2 n)\right] h^k_{v}(0)
  |\Lambda_b(v)\rangle &=& f^{1}_\Lambda \Psi_3^s(t_1,t_2) U(v)\,,
\nonumber\\
 \epsilon_{ijk}\langle 0| \left[u^i(t_1 n)C\gamma_5
 i  \sigma_{\bar n n} d^j(t_2 n)\right] h^k_{v}(0)
  |\Lambda_b(v)\rangle &=&  f^{3}_\Lambda \Psi_3^\sigma(t_1,t_2) U(v)
  \,,
\end{eqnarray}
where
\begin{eqnarray}
  \Psi(t_1,t_2) &=& \int_0^\infty \!\!d\omega_1  \int_0^\infty\!\! d\omega_2 \, e^{-it_1\omega_1 -it_2\omega_2} \psi(\omega_1,\omega_2)
\nonumber\\
&=&  \int_0^\infty\! \!\omega\, d\omega \int_0^1 du \,
e^{-i\omega(t_1u +it_2\bar u)} \widetilde\psi(\omega,u)\,, \\
\widetilde\psi_2(\omega,u) &=& \omega^2 u(1-u) \left[
\frac{1}{\varepsilon_0^4}e^{-\omega/\varepsilon_0} +
a_2C_2^{3/2}(2u-1)  \frac{1}{\varepsilon_1^4}
e^{-\omega/\varepsilon_1} \right] \, , \nonumber \\
 \widetilde\psi^s_3(\omega,u) &=& \displaystyle{\frac{\omega}{2\varepsilon_3^3}}\,e
 ^{-\omega/\varepsilon_3}\,, \nonumber \\
\qquad \widetilde\psi^\sigma_3(\omega,u) &=&
\frac{\omega}{2\varepsilon_3^3}(2u-1)\,e ^{-\omega/\varepsilon_3}\,
,
\end{eqnarray}
  $v_\mu= (n_\mu+\bar n_\mu)/2$, $v\cdot n=1$, $n\cdot \bar n =2$,  $
\widetilde\psi(\omega,u) =  \psi(u \omega,\bar u \omega)$, and $\bar
u = 1-u$. The $\widetilde{\psi}$ denotes the light-cone distribution
amplitudes $\widetilde{\psi}_2$, $\widetilde{\psi}_3^s$ and
$\widetilde{\psi}_3^\sigma$. The $\omega_1$ and $\omega_2$ are the
energies of the $u$ and $d$ quarks respectively, and
$\omega=\omega_1+\omega_2$, $\omega_1 = u \omega$ and $\omega_2 =
\bar u \omega$. The  $\varepsilon_0$, $\varepsilon_1$, $a_2$ and
$\varepsilon_3$ are nonperturbative parameters.

Substituting the above $c$ quark propagator and the corresponding
$\Lambda_b$ baryon  light-cone distribution amplitudes into the
correlation functions $\Pi^i_\mu(p,q)$, and completing the integrals
over the variables $x$ and $\lambda$, finally we obtain the
representation at the level of quark-gluon degrees of freedom,
\begin{eqnarray}
\Pi^i_{\mu}(p,q)&=& f_\Lambda^i\frac{1+\not\!v'}{2} \gamma_\mu U(v)
\int_0^{\infty}\omega'd\omega'\int_0^1 du\frac{1}
  {\omega \omega'-p\cdot v'}   \widetilde{\psi}(\omega',u)  + \cdots\, ,
 \end{eqnarray}

 After matching  with the hadronic representation below the continuum threshold $s_0$, we
obtain three  sum rules for the Isgur-Wise function $\xi(\omega)$,
\begin{eqnarray}
\xi(\omega)\exp\left[-\frac{\bar{\Lambda}}{T}\right] &=&
\int_0^{s_0}ds \int_0^1 du
\frac{s}{\omega^2}\widetilde\psi\left(\frac{s}{\omega},u\right)\exp\left[-\frac{s}{T}\right]\,,
\end{eqnarray}
where the $T$ is the Borel parameter, the  $\widetilde{\psi}$
denotes the $\widetilde{\psi}_2$, $\widetilde{\psi}_3^s$ and
$\widetilde{\psi}_3^\sigma$, thereafter we will denote the
corresponding sum rule as SRI, SRII, and SRIII  respectively. The
present sum rules do not suffer from uncertainties which originate
from the hadronic parameters $f_\Lambda^i$ as they cancel out
between the left side and the right side. In the light-cone QCD sum
rules, the hadronic parameters always make large contributions to
the uncertainties.  The four-particle light-cone distribution
amplitudes of the $\Lambda_b$ baryon are unknown, we only take into
account the contributions from the three-quark light-cone
distribution amplitudes. In case of the nucleon, the contributions
proportional to the gluon $G_{\mu\nu}$ can give rise to
four-particle (and five-particle) nucleon distribution amplitudes
with a gluon (or quark-antiquark pair) in addition to the three
valence quarks, their corrections are usually not expected to play
any significant roles \cite{DFJK}.

\section{Numerical result and discussion}
The input parameters  are taken as $\bar{\Lambda}=0.8\,\rm{GeV}$,
$s_0=1.2\,\rm{GeV}$, $T=(0.4-0.8)\,\rm{GeV}$, $\varepsilon_0=
200^{+130}_{-60}$~MeV, $\varepsilon_1= 650^{+650}_{-300}$~MeV,
$a_2=0.333^{+0.250}_{-0.333}$, and $\varepsilon_3 = 230$~MeV at the
energy scale $\mu=1\,\rm{GeV}$ \cite{Ball2008}.

The   nonperturbatuive parameters $\varepsilon_0$, $\varepsilon_1$,
$\varepsilon_3$, $a_2$  in the light-cone distribution amplitudes
are estimated by calculating the first few moments with the
 two-point QCD sum rules, the uncertainties are very large. In
numerical calculation, we observe that the uncertainties originate
from the nonperturbatuive parameters ($\varepsilon_0$,
$\varepsilon_1$, $\varepsilon_3$)   are  out of control.  In this
article, we take the central values and make a crude estimation.

The values of the  threshold parameter and Borel parameter
$s_0=1.2\,\rm{GeV}$ and $T=(0.4-0.8)\,\rm{GeV}$ are determined by
the  two-point QCD sum rules. The physical region of the Isgur-Wise
function $\xi (\omega)$ lies in the range $\omega=1-1.43$.  In
Fig.1, we plot the Isgur-Wise function $\xi (\omega)$  from the SRI
and SRII, respectively. From the figure, we can see that the
Isgur-Wise function $\xi (\omega)$ from the SRI is more stable with
variation of the Borel parameter $T$. At the interval
$T=(0.6-0.8)\,\rm{GeV}$, the curves of the $\xi (\omega)$ are more
flat than that of the $T=(0.4-0.6)\,\rm{GeV}$ and the predictions
are more robust, we can take the value $T=(0.6-0.8)\,\rm{GeV}$.
Finally we obtain the values,
\begin{eqnarray}
\xi(1)&=&1.09\pm0.05 \, , \nonumber\\
\xi(1)&=&1.30 \pm0.08\, ,
\end{eqnarray}
for  the SRI and SRII, respectively. Here only the uncertainties
originate from the Borel parameter $T$ are taken into account.
Although the $\xi(1)$ deviates from the normalization condition
$\xi(1)=1$, the central value $\xi(1)=1.09$ is rather good. From
Eq.(9), we can see that the model light-cone distribution amplitudes
are simple, more complicated distribution amplitudes maybe improve
the predictions. Furthermore, we have neglected the contributions
from the four-particle light-cone distribution amplitudes, their
contributions maybe large enough to smear the discrepancy. As the
four-particle light-cone distribution amplitudes of the $\Lambda_b$
baryon are unknown, we can not take into account their
contributions.

Taking  the exponential parametrization
$\xi(\omega)=\exp\left[-\rho^2(\omega-1)\right]$ and the
normalization condition $\xi(1)=1$, we  obtain the values of the
 slope parameter $\rho^2=1.10$ and  $\rho^2=0.85$ for the SRI and SRII respectively, which
are consistent with the estimation $\rho^2=1.35\pm0.12$ by  the QCD
sum rules  \cite{HuangMQ05},  the QCD sum rules also support much
smaller value $\rho^2=0.55\pm0.15$ \cite{Dai1996}; the present
prediction is rather good with the simple model.

The SRIII involves the  distribution amplitude
$\widetilde{\psi}_3^\sigma$, the integral $\int_0^1 du
\widetilde{\psi}_3^\sigma(\omega',u)=0$, so that $\xi(\omega)=0$,
the prediction is very poor.

\begin{figure}
\centering
  \includegraphics[totalheight=6cm,width=7cm]{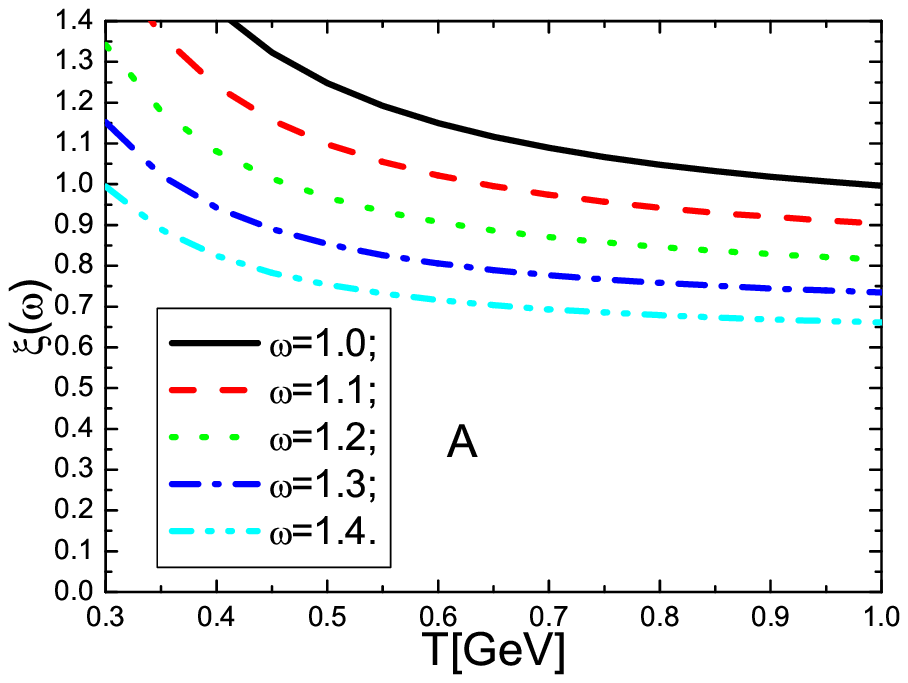}
 \includegraphics[totalheight=6cm,width=7cm]{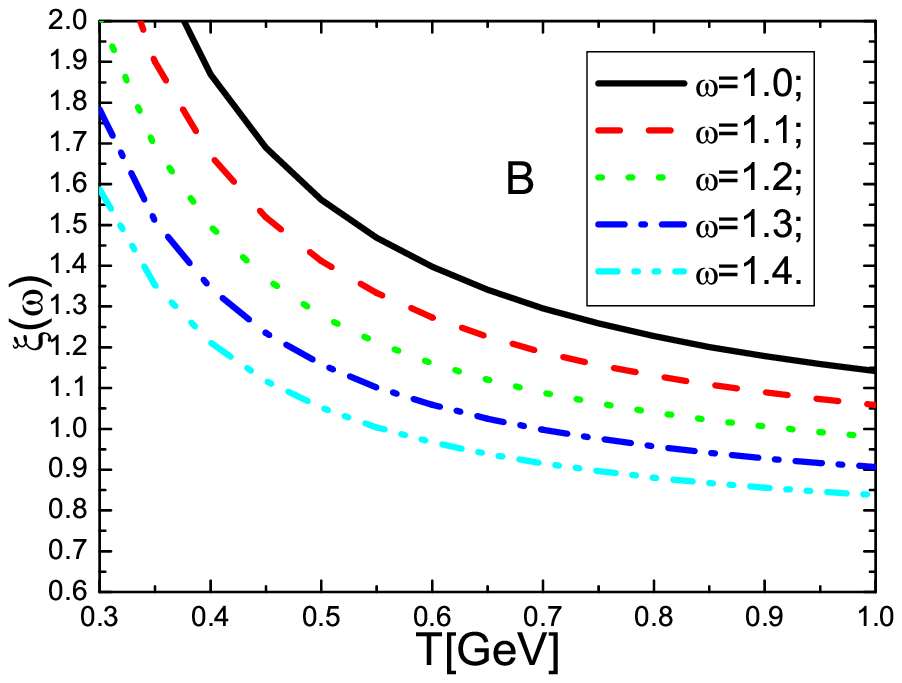}
       \caption{The Isgur-Wise function $\xi (\omega)$ with variation of the Borel parameter  $T$. The $A$ and $B$
       denote the   values from the SRI and SRII, respectively. }
\end{figure}

In Ref.\cite{Ball2008}, Ball et al observe   that the evolution
effects drive the light-cone distribution amplitudes to generate a
radiative tail  that falls off as $\ln(\omega_1/\mu)/\omega_1$ or
$\ln(\omega_2/\mu)/\omega_2$ at large energies, which is analogous
to the evolution behavior of the  $B$-meson light-cone distribution
amplitude \cite{Lange2003}. In this article, we obtain the sum rules
without the radiative $\mathcal {O}(\alpha_s)$ corrections, the
ultraviolet behavior of the $\widetilde{\psi}$ plays no role at the
leading order. Furthermore, the duality thresholds in the sum rules
are well below the region where the effect of the tail becomes
noticeable.

\section{Conclusion}
In this article, we use the light-cone QCD sum rules to relate the
$\Lambda_b$ baryon light-cone distribution amplitudes to the
Isgur-Wise function $\xi(\omega)$ of the $\Lambda_b \to \Lambda_c$
transition, and obtain a simple relation. The numerical value of the
Isgur-Wise function $\xi(\omega)$ is consistent with the prediction
of the  QCD sum rules. If the four-particle light-cone distribution
amplitudes are taken into account and the three-quark light-cone
distribution amplitudes are improved,  the prediction maybe better.

\section*{Acknowledgements}
This  work is supported by National Natural Science Foundation,
Grant Number  10775051, and Program for New Century Excellent
Talents in University, Grant Number NCET-07-0282.

\end{document}